\documentclass[times, 12pt]{article}

\usepackage{moreverb}

\newcommand\BibTeX{{\rmfamily B\kern-.05em \textsc{i\kern-.025em b}\kern-.08em
T\kern-.1667em\lower.7ex\hbox{E}\kern-.125emX}}

\usepackage{amssymb, color}
\usepackage{amsmath, epsfig, graphics, subfigure}

\newcommand{\be}{\begin{equation}}
\newcommand{\en}{\end{equation}}

\renewcommand{\vec}[1]{\boldsymbol{#1}}

\begin{document}

\title{\color{black}Slight compressibility
and sensitivity to changes in Poisson's ratio\color{black}}

\author{Michel Destrade$^{a,b}$, Michael D. Gilchrist$^b$, \\Julie Motherway$^b$, Jerry G. Murphy$^c$, \\[12pt]
$^a$School of Mathematics, Statistics and Applied Mathematics, \\
National University of Ireland Galway, Ireland; \\[4pt]
$^b$School of Mechanical and Materials Engineering, \\
University College Dublin, Belfield, Dublin 4, Ireland; \\[4pt]
$^c$Department of Mechanical Engineering, \\
Dublin City University, Glasnevin, Dublin 9, Ireland.\\}

\date{}

\maketitle

\begin{abstract}
 
Finite Element simulations of rubbers and biological soft tissue usually assume that the material being deformed is slightly compressible. 
It is shown here that in shearing deformations the corresponding normal stress distribution can exhibit extreme sensitivity to changes in Poisson's ratio. 
These changes can even lead to a reversal of the usual Poynting effect. Therefore the usual practice of arbitrarily choosing a value of Poisson's ratio when numerically modelling rubbers and soft tissue will, almost certainly, lead to a significant difference between the simulated and actual normal stresses in a sheared block because of the difference between the assumed and actual value of Poisson's ratio. The worrying conclusion is that simulations based on arbitrarily specifying Poisson's ratio close to $1/2$ cannot accurately predict the normal stress distribution even for the simplest of shearing deformations. It is shown analytically that this sensitivity is due to the small volume changes which inevitably accompany all deformations of rubber-like materials. To minimise these effects, great care should be exercised to accurately determine Poisson's ratio before simulations begin.

\end{abstract}

\newpage


 \section{Introduction }


When solid rubber-like materials are assumed to be incompressible so that only isochoric (i.e. volume preserving) deformations are allowed, then the theory of non-linear elasticity is known to be extremely effective in providing analytical predictions for their mechanical response when both the original geometry and boundary conditions are simple. These analytical solutions include shear, torsion, bending, straightening and inflation (see, e.g., Rivlin \cite{Rivl48}, Green and Zerna \cite{GrZe68}, or Ogden \cite{Ogde84}). 
To analyze more complex problems, use of Finite Element (FE) simulations is usually necessary. However, in most commercial finite element codes, incompressibility is not assumed \textit{ab initio} in order to prevent element locking (see, e.g., \cite{Malkus, DoBe99, BoWo08}). 
Thus in FE analyses, slight compressibility is usually assumed and the solution for fully incompressible materials is obtained in a limit process if necessary (see, e.g., \cite{BoWo08, SiTa91, SiHu97, Cris97}). Indeed, because all solid rubbers are to some extent compressible, models for slight compressibility have been investigated independently of any FE considerations (see, e.g., Ogden \cite{Ogde84}, Horgan and Murphy \cite{HoMu07, HoMu09a, HoMu09b, HoMu09c} and references cited therein). 

Rubber components are often subjected to shearing deformations in applications and biological soft tissue is often sheared \emph{in vivo} (see, for example, LeGrice \emph{et al.} \cite{Grice} for a discussion on shear strain in the left ventricular myocardium \color{black}or Horgan and Murphy \cite{HoMu11} on the simple shearing of tissues). \color{black}
It is therefore essential that accurate, reliable and efficient models be available to predict shear behaviour. In a recent paper, Gent \emph{et al.} \cite{Gent07} conducted some numerical simulations using the commercial code ABAQUS of the simplest possible shearing deformation, where one face of a block, modelled as a compressible neo-Hookean solid, is displaced relative to the parallel face. This shearing deformation will be called here \emph{experimental simple shear}. One set of their results is puzzling: in their Figures 12 and 13, an infinitesimal change in Poisson's ratio $(\nu)$ leads to a large change in the predicted values of the normal stress components, with, in some cases, a small change in Poisson's ratio leading to a change in the sign in the components, leading effectively to an `inverse' Poynting effect. These numerical results have been confirmed by our own numerical simulations (\S 2) for which the adopted procedure mirrors, as closely as possible, the earlier work. 

In \S 3 this extreme sensitivity of the normal stresses to changes in $\nu$ is explained by combining a realistic mathematical model of experimental simple shear that incorporates the possibility of small volume changes with the  constitutive models of slight compressibility used in FE simulations. It will be shown that the predicted normal stresses for the slightly compressible neo-Hookean material obtained from this combination display exactly the sensitivity to changes in the Poisson ratio as was first observed by Gent \emph{et al.} \cite{Gent07}. The shearing strains considered here are only moderate in range, with a maximum shear strain of 100\% being imposed. It is likely that the extreme sensitivity noted here is more pronounced for larger strains.

We suggest that the results discussed above have significant implications for the numerical simulation of materials traditionally considered as being incompressible. We note that in practice, when simulating these materials, experimental values of the corresponding  Poisson ratio are not known. Instead, most engineers adopt the pragmatic solution of specifying instead a value of Poisson's ratio `close' to $1/2$, assuming implicitly that  the specific value of Poisson's ratio adopted will not have a significant impact on the numerical results and, in particular, that small percentage changes in the value of Poisson's ratio result in corresponding small percentage changes in the stress distribution. Most simulations of biological soft tissue, for example, are conducted on this basis. However, the simulations of Gent \emph{et al.} \cite{Gent07} and our own numerical experiments (\S 2) all suggest that, with the models currently used, a small change in the value of Poisson's ratio chosen (say from 0.499 to 0.495) will result in significant changes in the stress distribution. There is almost certainly a difference between the assumed and actual values of Poisson's ratio and this should generally result in a significant difference between predicted and actual stress distributions. Our conclusion is that an accurate determination of Poisson's ratio is essential if FE simulations are to yield accurate normal stresses in shearing deformations
\color{black}
(See also Horgan and Murphy \cite{HoMu10} for a discussion of the results of Gent \emph{et al.} \cite{Gent07}  relevant to the present paper.) \color{black}

It is finally noted that the seemingly obvious solution to this extreme sensitivity to Poisson's ratio of simply simulating \emph{perfect} incompressibility is not a valid approach. No material is perfectly incompressible and assuming this idealisation (denoted in our results by $\nu=1/2$) will result in a difference between the idealised value assumed and the actual value of Poisson's ratio. This difference will lead to a significant percentage error in the predicted normal stresses.


 \section{Modelling slight compressibility}
 
 
 Many different constitutive models have been proposed to reflect slight deviations from incompressibility on assuming that the material is homogeneous, isotropic and hyperelastic. We refer to the recent papers \cite{HoMu07, HoMu09a, HoMu09b, HoMu09c}  for background, references to the pertinent literature and, very importantly, a summary of the fit of these models, and, in particular, the models used here, with experimental data. Here we briefly describe the usual form of such slightly compressible (or almost incompressible) strain-energy functions used in the FE simulations. 
 
Assume that an \textit{incompressible} strain-energy function $W=\psi \left(\lambda_{1} ,\lambda_{2} ,\lambda_{3} \right)$, where $\psi $ is a symmetric function of the principal stretches $\lambda_{i} $, has been obtained that gives an excellent fit to the experimental data collected from some set of material characterization experiments. 
Then the corresponding \emph{slightly compressible} form implemented in FE models is usually of the form
\begin{equation} \label{GrindEQ__2_1_} 
W =\psi \left(\bar{\lambda }_{1} , \bar{\lambda }_{2} , \bar{\lambda }_{3} \right) + F\left(i_{3} \right),  
\end{equation} 
 where 
\begin{equation} \label{GrindEQ__2_2_} 
i_{3} \equiv \lambda _{1} \lambda _{2} \lambda _{3},\qquad
\bar{\lambda}_{i} \equiv \lambda _{i} /i_{3}^{1/3},                               
\end{equation} 
$F$ is some specified polynomial function of its indicated argument, and the $\bar \lambda$'s are the so-called `deviatoric stretches' (see the ABAQUS \cite{abaq} and ANSYS \cite{ansys} manuals).

On assuming that $\psi \left(1,1,1\right)=0$ to ensure zero strain-energy in the reference configuration for the incompressible material, it follows from \eqref{GrindEQ__2_1_} that we need to assume that 
\be \label{F(1)}
F(1)=0,
\en 
for the same property to hold for the slightly compressible material. 
To ensure zero stress in the reference configuration, it is sufficient to require that
\begin{equation} \label{F'(1)} 
F'(1)=0.                                                           
\end{equation} 
Finally, to ensure compatibility with the classical linear form of the elastic strain-energy function on restriction to infinitesimal deformations, it is required that
\begin{equation} \label{F''(1)} 
\frac{\partial ^{2} \psi }{\partial \bar{\lambda }_{1}^{2} } \left(1,1,1\right) 
-\frac{\partial ^{2} \psi }{\partial \bar{\lambda }_{1} \partial \bar{\lambda }_{2} }\left(1,1,1\right) 
+\frac{\partial \psi }{\partial \bar{\lambda }_{1}}\left(1,1,1\right)
= 2\mu, 
\qquad
F''(1) = \kappa,   
\end{equation} 
where $\mu$ and $\kappa$ are the infinitesimal shear and bulk moduli, respectively.
Recall that they are connected to Poisson's ratio $\nu$ through
\begin{equation}
\label{connect}
 \kappa / \mu =2 (1+\nu)/[3(1-2\nu)].
 \end{equation}

Motivated by these relations, it is now assumed that the $\psi $ term in \eqref{GrindEQ__2_1_} is proportional to $\mu $ and that \textit{F} is proportional to $\kappa$.
 On setting $i_{3} =1$ in \eqref{GrindEQ__2_1_}, one recovers the original strain-energy function for the corresponding incompressible material. 
It is important to emphasise that the main motivation for strain-energy functions of the form \eqref{GrindEQ__2_1_} seems to be mathematical convenience: a simple additive term in $i_3$ is included to account for the compressibility so that, on restriction to infinitesimal deformations, $\psi$ and \textit{F} are proportional to the shear modulus $\mu$ and the bulk modulus $\kappa$, respectively (see e.g., Horgan and Murphy \cite{HoMu09c} for further details). 

Many FE codes are based on models of the form \eqref{GrindEQ__2_1_}.
The initial moduli are usually assumed to be such that $\kappa/ \mu \gg 1$ in order to model slight incompressibility. 
In the variational formulation, the bulk modulus $\kappa$ is called the penalty number (see, e.g., \cite{Malkus, JiLe94, BoWo08}) and is chosen to be large. From \eqref{connect}, we find that for $\nu = 0.49, 0.499, 0.4999$,  we have $\kappa/ \mu \simeq 50, 500, 5000$, respectively. 
\color{black}
The default setting in Abaqus/Explicit is quite low, as it assumes that $\kappa/\mu = 20$, corresponding to a Poisson ratio of $\nu \simeq 0.475$.
\color{black}

Equivalently, the model can be written in terms of the Cauchy-Green strain invariants, defined in terms of the principal stretches $\lambda_{i}$, $i = 1, 2, 3$, as follows,
\begin{equation} \label{GrindEQ__2_5_} 
I_{1} =\lambda _{1}^{2} + \lambda_{2}^{2} + \lambda _{3}^{2} ,\qquad 
I_{2} =\lambda _{1}^{2} \lambda _{2}^{2} +\lambda _{1}^{2} \lambda _{3}^{2} +\lambda _{2}^{2} \lambda _{3}^{2} ,\qquad 
I_{3} \equiv i_{3}^{2} =\lambda _{1}^{2} \lambda _{2}^{2} \lambda _{3}^{2} .                    
\end{equation} 
The corresponding strain-energy function that is used in FE simulations to model almost incompressible behavior then has the equivalent form
\begin{equation} \label{GrindEQ__2_6_} 
W =\psi \left(\bar{I}_{1} ,\bar{I}_{2} \right) + F(I_{3}^{1/2}),\qquad \text{where} \qquad
\bar{I}_{1} =I_{1} / I_{3}^{1/3},\qquad 
\bar{I}_{2} = I_{2} /I_{3}^{2/3} .                             
\end{equation} 
Further details can be found, for example, in Crisfield \cite{Cris97}.

The general representation of the stress-strain relations for compressible, non-linear elasticity can be found, for example, in Atkin and Fox \cite{AkFo80} and is given by
\be
\vec{T} = \alpha_{0} \vec{I} + \alpha_{1} \vec{B} + \alpha_{2} \vec{B}^2,                                                    
\label{GrindEQ__2_7_} 
\en
where
\begin{equation} \label{GrindEQ__2_8_} 
\alpha _{0} =2I_{3}^{{1 \mathord{\left/{\vphantom{1 2}}\right.\kern-\nulldelimiterspace} 2} } \frac{\partial W}{\partial I_{3} } ,\qquad 
\alpha _{1} =\frac{2}{I_{3}^{{1 \mathord{\left/{\vphantom{1 2}}\right.\kern-\nulldelimiterspace} 2} } } \left(\frac{\partial W}{\partial I_{1} } +I_{1} \frac{\partial W}{\partial I_{2} } \right),\qquad 
\alpha _{2} =-\frac{2}{I_{3}^{{1 \mathord{\left/{\vphantom{1 2}}\right.\kern-\nulldelimiterspace} 2} } } \frac{\partial W}{\partial I_{2} }.                        
\end{equation} 
Here $\vec{T}$ is the Cauchy stress, $\vec{I}$ is the identity tensor, and $\vec{B}$ is the left Cauchy-Green strain tensor, defined in terms of the deformation gradient tensor $\vec{F}$ through the relation $\vec{B} = \vec{FF}^{T}$. 
The stress-strain relation for the almost incompressible material, \eqref{GrindEQ__2_6_}, therefore has the form \eqref{GrindEQ__2_7_} with the following coefficients:
\begin{align} \label{GrindEQ__2_9_} 
& \alpha _{0} =-\frac{2}{3I_{3}^{1/2}}  \bar{I}_{1} \frac{\partial \psi }{\partial \bar{I}_{1} } - \frac{4}{3I_{3}^{1/2}} \bar{I}_{2} \frac{\partial \psi }{\partial \bar{I}_{2} } +F'(I_{3}^{1/2}), \notag \\[4pt]
& \alpha _{1} =\frac{2}{I_{3}^{5/6}} \left(\frac{\partial \psi }{\partial \bar{I}_{1} } +\bar{I}_{1} \frac{\partial \psi }{\partial \bar{I}_{2} } \right),\notag \\[4pt]
& \alpha _{2} =-\frac{2}{I_{3}^{7/6}} \frac{\partial \psi }{\partial \bar{I}_{2}}.    
\end{align} 

The stress-strain relation for almost incompressible materials can thus be put in the form
\be
\vec{T} = F'(i_3) \vec{I} + \hat{\alpha }_{0} \vec{I} + \alpha _{1} \vec{B} + \alpha_{2} \vec{B}^2,  
\label{general}
\en
with $\alpha _{1}$, $\alpha _{2}$ given in \eqref{GrindEQ__2_9_}$_{2,3}$ and $\hat{\alpha }_{0} = \alpha_0 - F'(i_3)$, i.e.
\be
\hat{\alpha }_{0} = -\frac{2}{3I_{3}^{1/2}}  \bar{I}_{1} \frac{\partial \psi }{\partial \bar{I}_{1} } - \frac{4}{3I_{3}^{1/2}} \bar{I}_{2} \frac{\partial \psi }{\partial \bar{I}_{2} }.
\en
Thus the stress is always the sum of two parts: a hydrostatic term proportional to the volumetric function $F (i_3)$, and therefore of the order of the infinitesimal bulk modulus $\kappa$, and another term of the order of the infinitesimal shear modulus $\mu$.
 \begin{figure}
\center
	 \subfigure{\epsfig{figure=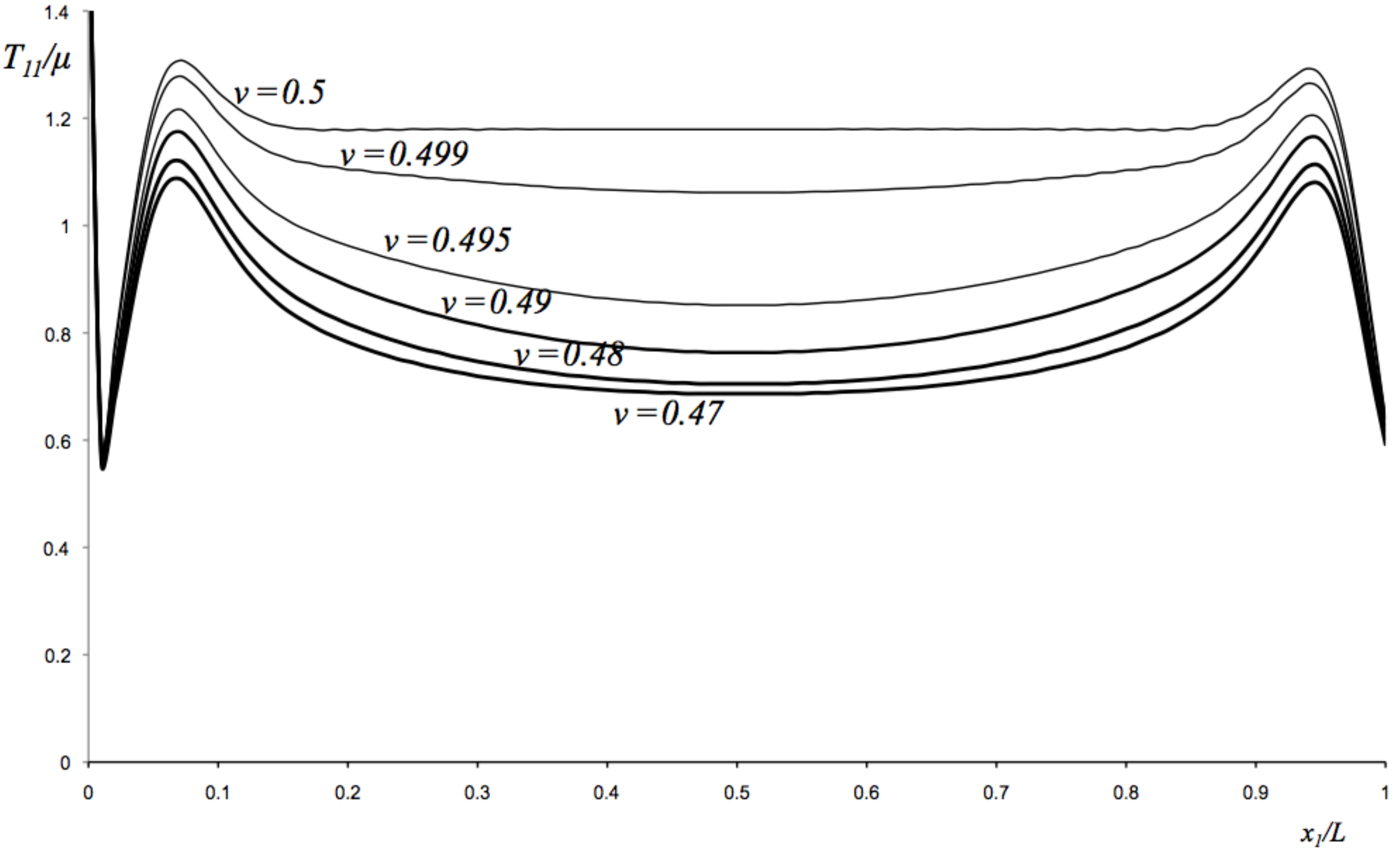,width=.49\textwidth}}
\,
 \subfigure{\epsfig{figure=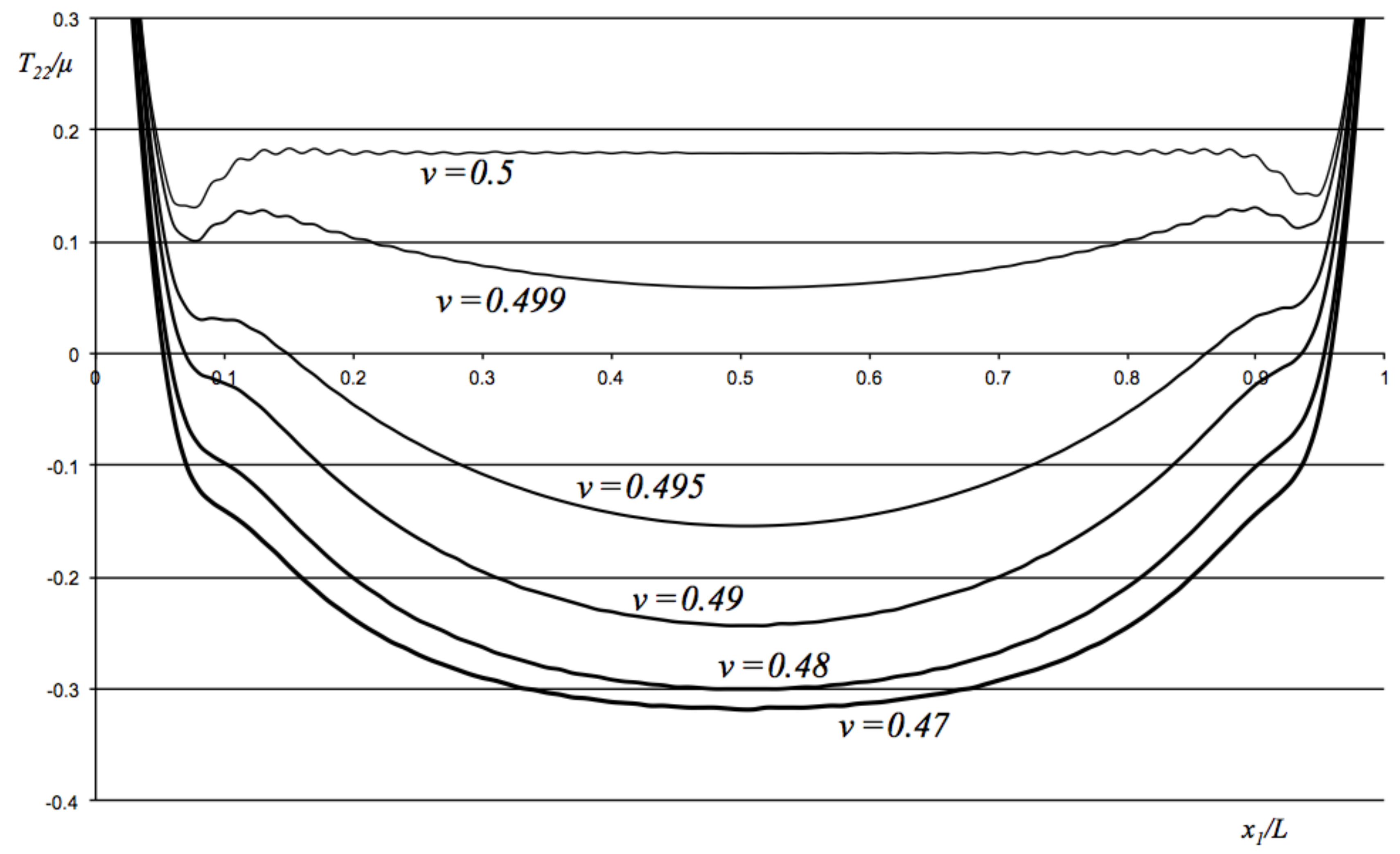,width=.49\textwidth}}
	\caption{Normal stress variations along the horizontal center line of a sheared block of a slightly compressible neo-Hookean material (amount of shear: 1, block's aspect ratio: 10). 
	The Cauchy stress components $T_{11}$ (left) and $T_{22}$ (right) are highly sensitive to small changes in the values of Poisson's ratio $\nu = 0.5, 0.499, 0.495, 0.49, 0.48, 0.47$. }
	\label{fig:gent}
\end{figure}

The numerical simulations of Gent \emph{et al.} \cite{Gent07} are based on the slightly compressible neo-Hookean specification of this strain-energy function, readily implemented in ABAQUS \cite{abaq}. 
It has the form
\begin{equation} \label{GrindEQ__2_11_} 
W =\frac{\mu }{2} \left(\bar{I}_{1} -3\right)+F\left(i_{3} \right)=\frac{\mu }{2} \left(\bar{\lambda }_{1}^{2} +\bar{\lambda }_{2}^{2} +\bar{\lambda }_{3}^{2} -3\right)+F\left(i_{3} \right),                              
\end{equation} 
where the constitutive volumetric function \textit{F} is the quadratic form 
\begin{equation} \label{GrindEQ__2_12_} 
F(i_3) = \frac{\kappa}{2} \left(i_{3} -1\right)^{2},                                                 
\end{equation} 
which is the simplest form to satisfy \eqref{F(1)}, \eqref{F'(1)}, and \eqref{F''(1)}$_2$. For this material the constitutive coefficients simplify to
\begin{equation}
\hat{\alpha }_{0} = - \dfrac{\mu I_1}{3I_{3}^{5/6}}, \qquad 
\alpha_1 =  \dfrac{\mu}{I_{3}^{5/6}}, \qquad
\alpha_2 = 0,
\en
and the following stress-strain relation is obtained:
\be \label{neo}
\vec T = \kappa (I_3^{1/2} -1)\vec I + \dfrac{\mu}{I_3^{5/6}}\left(\vec B - \tfrac{1}{3}I_1 \vec I\right).
\en
%
Predictions based on the neo-Hookean model have been found to yield a good fit with data from a variety of experiments on rubber-like materials for small to moderate strains. Since moderate strains are only considered here, the results obtained in our analyses are therefore likely to be obtained from simulations based on other forms of the deviatoric part of the strain-energy function.

The form \eqref{GrindEQ__2_11_} was assumed in our repeating, and furthering, of the numerical simulations of Gent \emph{et al.} \cite{Gent07}. We entered their physical and geometrical parameters into ABAQUS 6.7 to try to replicate their results for a sheared neo-Hoohean block. We obtained the stresses and strains throughout the thickness. 
We varied the mesh resolution, the time step duration and the number of increments used and found virtually no difference with their simulations.
Figure \ref{fig:FEM0_5} shows an example of the stress field distribution in the sheared block, displaying clearly an almost homogeneous field except near the edges.
Further pictures obtained for the stress and strain distributions are identical to those of Gent \emph{et al.} \cite{Gent07}, see Figure \ref{fig:gent}. 
\begin{figure}
\centering
 \epsfig{figure=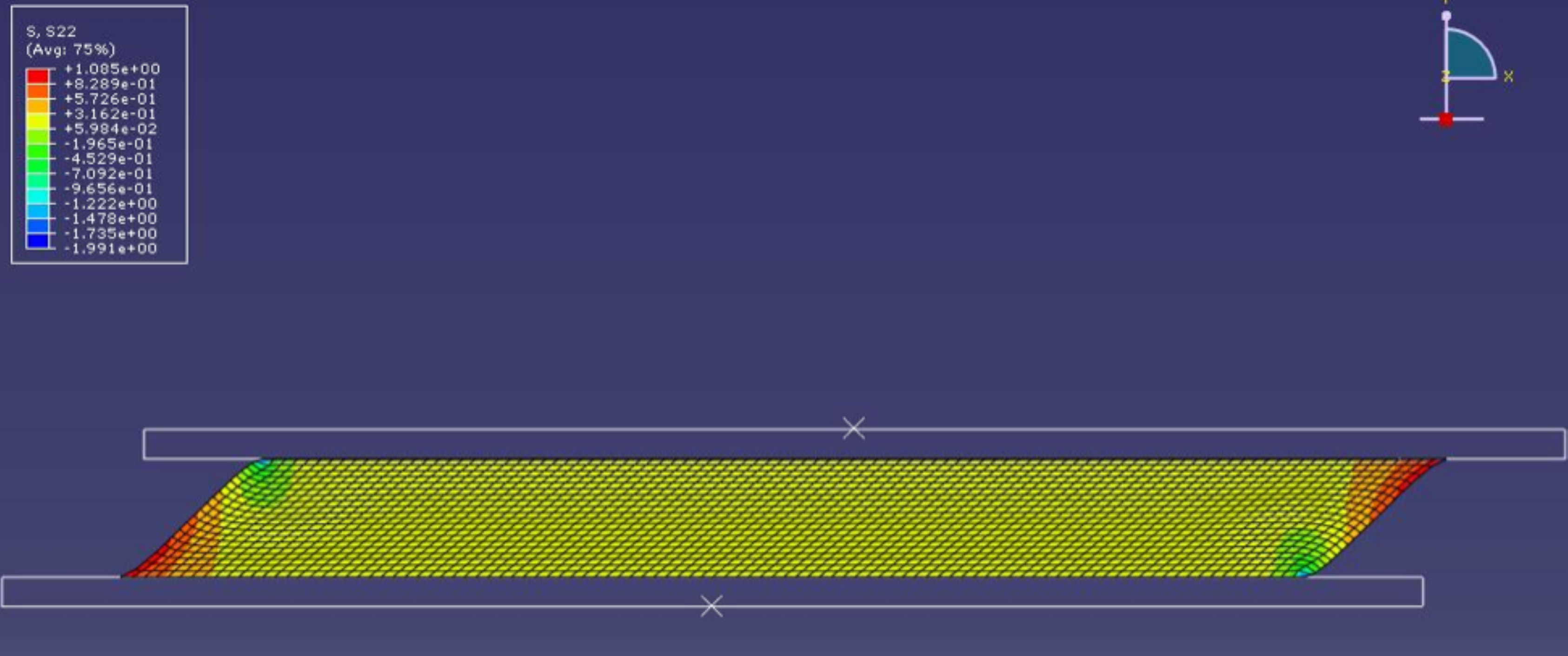, width=.9\textwidth}
 \caption{Finite element simulation of the simple shear of a block with the same dimensions and characteristics as the block of Gent et al. \cite{Gent07} (amount of shear: 1, block's aspect ratio: 10).
 Here Poisson's ratio is $\nu = 0.5$ (perfect incompressibility) and the normal stress $T_{22}$ is tensile throughout the block, except in small areas near the compressed corners.}
 \label{fig:FEM0_5}
\end{figure}

We also carried out two re-meshes in addition to the original mesh \cite{Gent07} of 20 (thickness) by 100 (length),  i.e. 2,000 rectangular elements. 
The first re-mesh was a square mesh of $20 \times 200$ (i.e. 4,000 elements) and the second re-mesh was a square mesh of $40 \times 400$ (i.e. 16,000 elements). 
Each new mesh was run for the two following values of Poisson's ratio: $\nu = 0.49$ and $\nu= 0.5$. 
These re-meshes had no effect on the original results of Gent \emph{et al.} \cite{Gent07}.
Similarly, we varied the time step and incrementation, from their original \cite{Gent07} values of time step = 4 seconds, 400 increments (giving a shearing rate of  2 mm/s). 
Thus we halved (time=2, increments=200, shearing rate=4mm/s) and then doubled (time=8, increments=800, shearing rate=1mm/s) the time steps and increments. These time changes also has no effect.

In conclusion, we found, just like Gent \emph{et al.} \cite{Gent07}, that for the compressible neo-Hookean model \eqref{GrindEQ__2_11_}-\eqref{GrindEQ__2_12_} infinitesimal changes in the value of Poisson's ratio can lead to finite changes in the simulated stress distribution. Our simulations suggest that this problem is especially acute for $\nu \in \left[0.495,0.5\right]$ and that that the problem of extreme sensitivity of the normal stresses to Poisson's ratio increases the closer the value of $\nu$ is to $1/2$. These features are evident in Figure \ref{fig:gent} and, perhaps more emphatically, in a comparison of Figures \ref{fig:FEM0_5} and  \ref{fig:FEM0_495}. In Figure \ref{fig:FEM0_5}, where perfect incompressibility has been assumed, the stress is almost homogeneous, except for more complicated behaviour near the corners, and tensile throughout a block with aspect ratio 1:10 sheared by a unity amount. Figure \ref{fig:FEM0_495} displays the normal stress component $T_{22}$ distribution with Poisson's ratio $\nu = 0.495$. In contrast with Figure \ref{fig:FEM0_5}, the stress is no longer homogenous: it can be seen there that $T_{22}$ is tensile and non-uniform near the slanted faces, but is compressive towards the centre of the specimen.

\begin{figure}
\centering
 \epsfig{figure=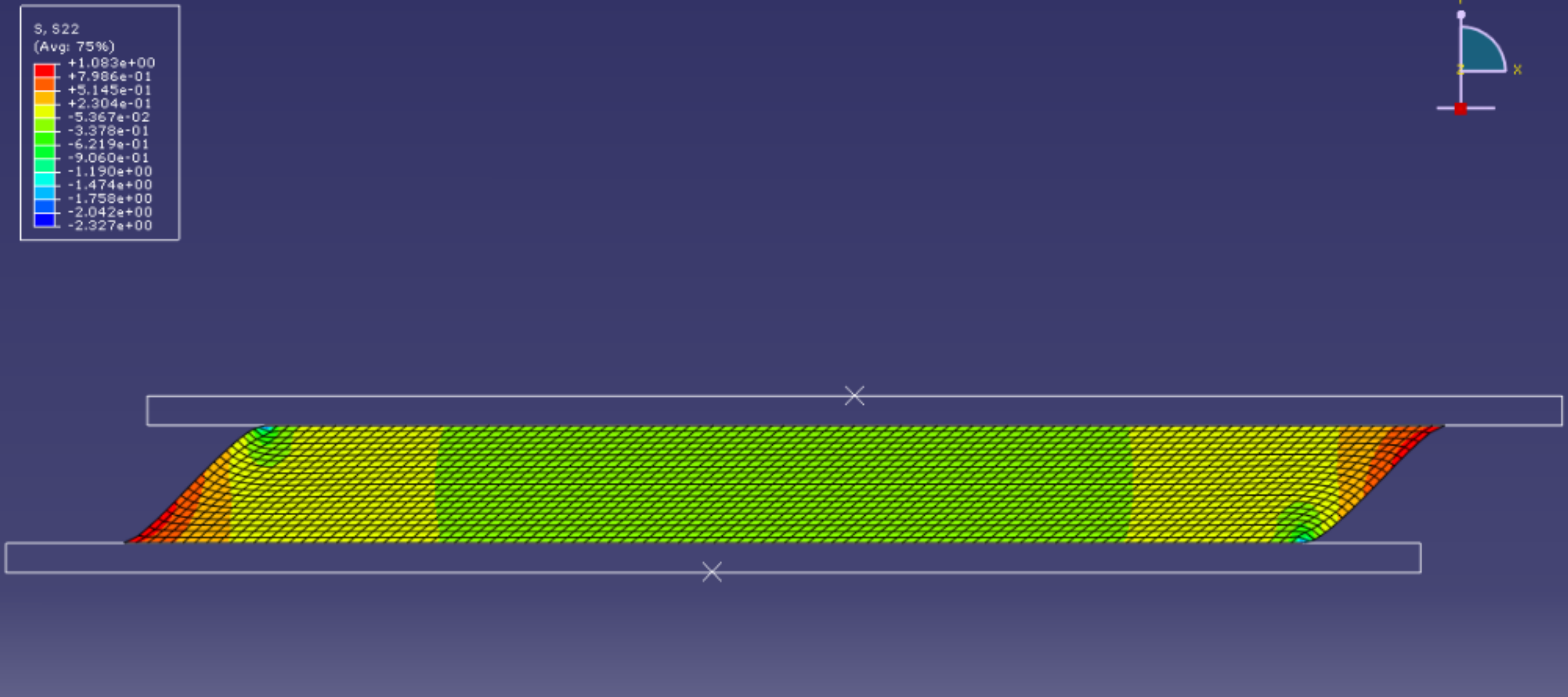, width=.9\textwidth}
 \caption{Finite element simulation of the simple shear of a block with the same dimensions and characteristics as the block of Gent et al. \cite{Gent07} (amount of shear: 1, block's aspect ratio: 10).
 Here Poisson's ratio is $\nu = 0.495$ (slight compressibility) and the normal stress $T_{22}$ is uniform and compressive in a large part of the center of the block.}
 \label{fig:FEM0_495}
\end{figure}

It is natural to see if the problems in simulating slightly compressible materials are reflected in analytical approaches to simple shear because their specific cause might then be isolated. To investigate these problems analytically, the classical mathematical approximation to experimental simple shear is generalised in the next section and the corresponding stress distribution obtained for both the general slightly compressible model \eqref{general} and for the slightly compressible neo-Hookean model \eqref{neo} used in FE simulations.

 
\section{Analytical approaches to simple shear}


 The classical mathematical model of simple shear assumes that a block of rubber is deformed according to
\begin{equation} \label{GrindEQ__3_1_} 
x_{1} =X_{1} + \gamma X_{2}, \qquad x_{2} =X_{2},\qquad x_{3} =X_{3} ,                                           
\end{equation} 
where $\left(X_{1}, X_{2}, X_{3} \right)$ and $\left(x_{1}, x_{2}, x_{3} \right)$ denote the Cartesian coordinates of a typical particle before and after deformation, respectively, and $\gamma$ is an arbitrary dimensionless constant called the \emph{amount of shear}. 
Within the context of nonlinear elasticity, this problem was first considered by Rivlin \cite{Rivl48} and has been much discussed and analyzed since. The reference works of Green and Zerna \cite{GrZe68}, Truesdell and Noll \cite{TrNo65}, and Ogden \cite{Ogde84}, for example, all consider various aspects of this problem. 

Intuitively it seems reasonable that this homogeneous deformation should accurately model experimental simple shear, especially within the bulk of the material and for modest amounts of shear.  The corresponding in-plane stress distribution for the general strain-energy function adopted in FE simulations to model almost incompressible rubbers is easily obtained from \eqref{GrindEQ__2_9_}, \eqref{general} and is given by
\begin{align} \label{GrindEQ__3_2_} 
& T_{11} =\hat{\alpha }_{0} +\alpha _{1} \left(1+\gamma^{2} \right)+\alpha _{2} \left(1+3 \gamma ^{2} + \gamma ^{4} \right), \notag \\
& T_{22} =\hat{\alpha }_{0} +\alpha _{1} +\alpha _{2} \left(1+  \gamma ^{2} \right), 
\notag \\
& T_{12} =\alpha _{1} \gamma +\alpha _{2} \gamma \left(2+ \gamma ^{2} \right).
\end{align} 

Modelling experimental simple shear using the intuitive deformation \eqref{GrindEQ__3_1_} therefore results in the in-plane stress field for almost incompressible materials of the form \eqref{GrindEQ__2_1_} being independent of the volumetric function $F$.  
This is not surprising given that classical simple shear is an isochoric deformation  ($i_3=1$). 
Therefore, for a fixed value of $\mu $, the in-plane stress field is \textit{independent} of the bulk modulus $\kappa$, or equivalently, of the Poisson ratio $\nu$. 
This is in conflict with the numerical predictions discussed earlier.

This conflict can be explained by the fact that using the isochoric approximation \eqref{GrindEQ__3_1_} to model experimental simple shear is too restrictive and does not capture an essential feature of every deformation of rubbers, i.e., that an infinitesimal volume change \emph{always} occurs and therefore occurs  in experimental simple shear in particular. 
To get an idea of the order of the change, we rely on the classical experimental data of Penn \cite{Penn70}. 
In a series of elegant experiments on peroxide vulcanizates of natural gum rubber using a dilatometer technique, Penn measured the volume change in simple \emph{tension} for axial stretches of the order of 2.  
The volume change, calculated as $i_{3} -1$, was found to be of the order of 0.0001. No comparable data exist for simple shear.
We now investigate the effect, if any, of a change in volume.

We model the infinitesimal volume change with the following simple generalization of the classical approach \eqref{GrindEQ__3_1_}, 
\begin{equation} \label{GrindEQ__3_3_} 
x_{1} = (1+\epsilon_1)X_{1} + \gamma (1+\epsilon_2) X_{2} ,\qquad 
x_{2} = \left(1+ \epsilon_2 \right)X_{2} ,\qquad 
x_{3} = \left(1+ \epsilon_3 \right)X_{3},\qquad 
\end{equation} 
where the $\epsilon$'s are infinitesimal quantities.
This deformation seems intuitively to capture the main features of the deformation of a block when one face is displaced relative to a parallel face. 
It corresponds to a simple shear of amount $\gamma$ superposed on a triaxial extension, where the $\epsilon$'s are the infinitesimal elongations.

Retaining only the first-order terms in the $\epsilon$'s gives $i_3 \simeq 1 + \epsilon$, where $\epsilon \equiv \epsilon_1 + \epsilon_2 + \epsilon_3$ is the volume change.
It follows that, to the first-order in $\epsilon $, the constitutive function $F'(i_{3})$ occurring in the stress-strain relation \eqref{general} for almost incompressible materials can be approximated as
\begin{equation} \label{GrindEQ__3_7_} 
F'(i_{3}) \simeq \kappa \epsilon,                                                        
\end{equation} 
where we used \eqref{F'(1)} and \eqref{F''(1)}$_2$.
	
The corresponding in-plane stresses are given by Rajagopal and Wineman \cite{RaWi87} as 
\begin{align}\label{GrindEQ__3_8_} 
& T_{11} = \kappa \epsilon + \hat{\alpha}_{0} + \alpha _{1} (\lambda_1^2 + \gamma^2 \lambda_2^2) + \alpha_2 [(\lambda_1^2 + \gamma^2\lambda_2^2)^2 + \gamma^2\lambda_2^4],\notag \\ 
& T_{22} = \kappa \epsilon + \hat{\alpha}_{0} + \alpha_1\lambda_2^2 + \alpha_2 (1 + \gamma^2)\lambda_2^4, \notag \\
& T_{12} = \alpha _{1} \gamma \lambda_2^2 + \alpha_2 \gamma \lambda_2^2[\lambda_1^2 + (1+\gamma^2)\lambda_2^2],
\end{align} 
where $\lambda_i \equiv 1+\epsilon_i$ are the stretch ratios (here it is understood that their powers are linearized with respect to the $\epsilon$'s).
Compare these general expressions with those for simple shear given in \eqref{GrindEQ__3_2_}:
the shear stress $T_{12}$ is again independent of the bulk modulus and volume change, but now both normal stresses $T_{11}$ and $T_{22}$ contain a new $\kappa \epsilon $ term. 
This is  in agreement with the effects observed in the numerical simulations, showing that the normal stresses, not the shear stress, are sensitive to volume changes (see Figures \ref{fig:gent} and \ref{fig:shear-i_3}(a)).

 \begin{figure}
\center
	 \subfigure{\epsfig{figure=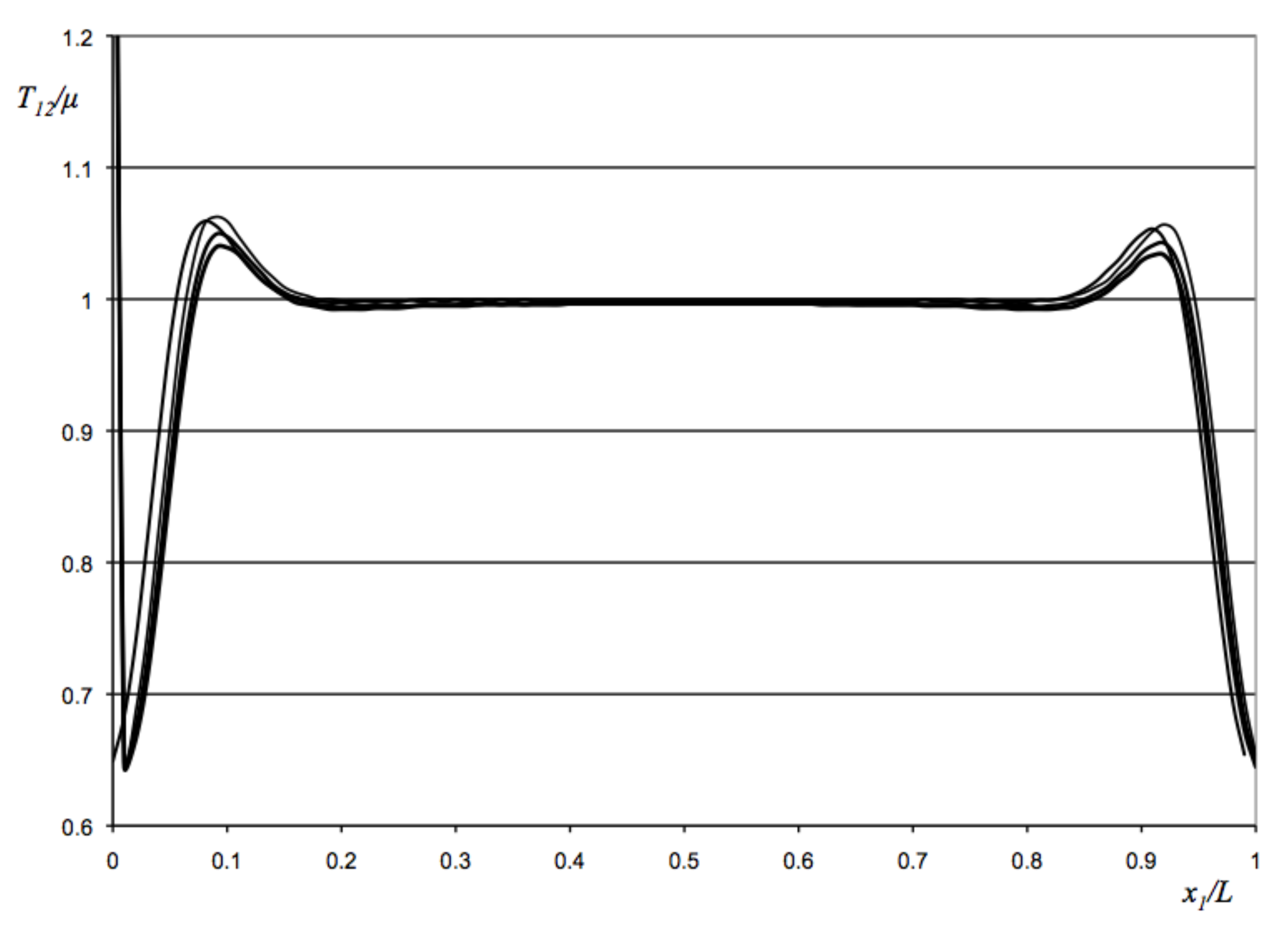,width=.49\textwidth}}
\,
 \subfigure{\epsfig{figure=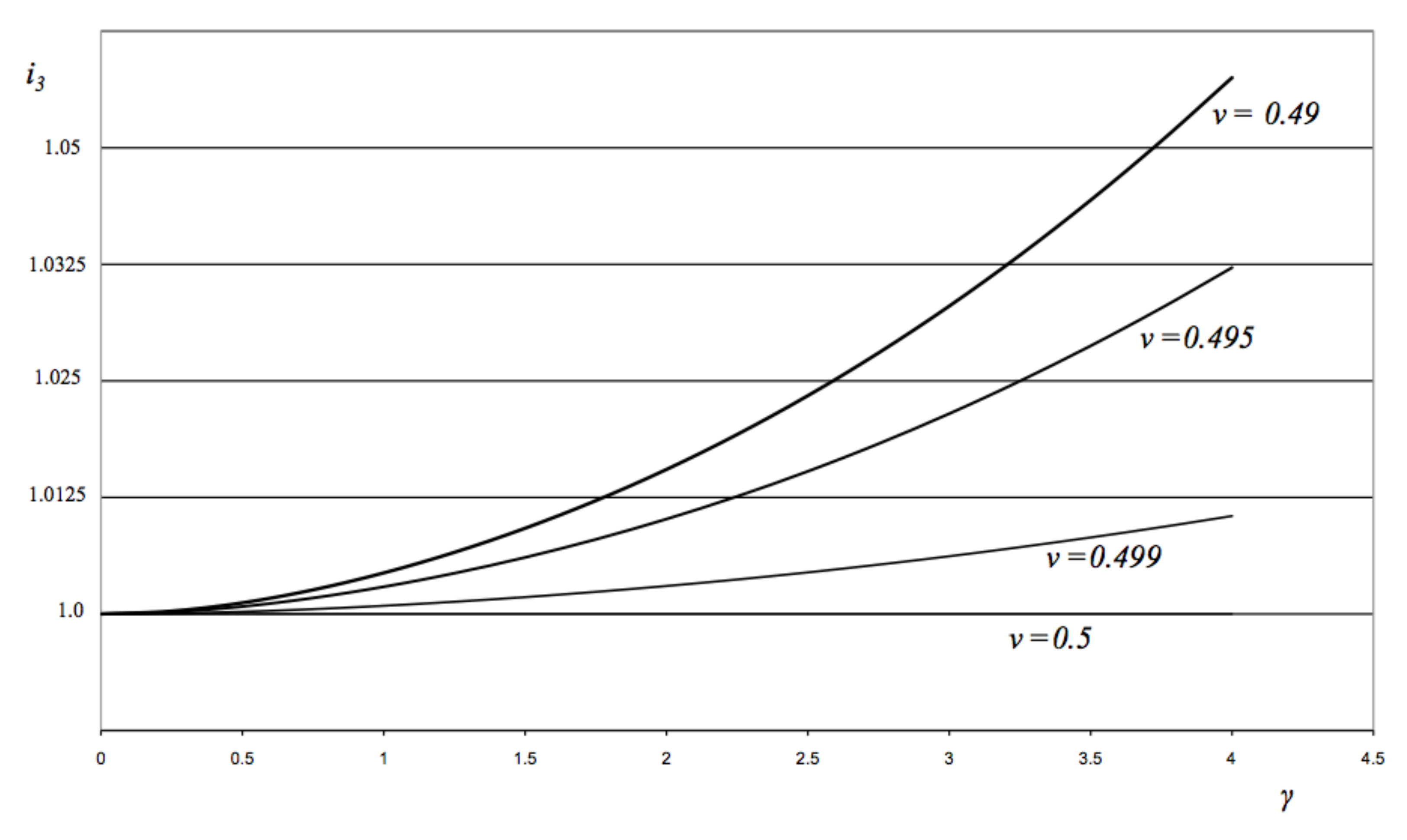,width=.49\textwidth}}
	\caption{On the left: Shear stress variation along the horizontal center line of a sheared block of a slightly compressible neo-Hookean material (amount of shear: 1, block's aspect ratio: 10). 
	The Cauchy stress component $T_{12}$ is not very sensitive to small changes in the values of Poisson's ratio $\nu = 0.5, 0.499, 0.495, 0.49$. 
	On the right: Volume change during the large shear of a slightly compressible block of a neo-Hookean material with an aspect ratio of 10.}
	\label{fig:shear-i_3}
\end{figure}

Now we try to mirror the numerical simulations: we specialize the stress-strain relation to that of the slightly compressible neo-Hookean solid \eqref{neo}; we impose a parallel displacement for the moving face with respect to the fixed face (so that $\lambda_2=1$); and we impose a two-dimensional formulation (so that $\lambda_3=1$). 
From  \cite{RaWi87} we find that the first principal invariant is $I_1 = 3 + \gamma^2 + 2\epsilon$, where we linearized with respect to $\epsilon = \epsilon_1$. For this specialisation the normal stress component $T_{22}$, for example, is easily computed and we find that 
\be \label{negative}
\frac{T_{22}}{\mu} = -\dfrac{\gamma^2}{3} + \left(\dfrac{2\nu}{1-2\nu}+\dfrac{5\gamma^2}{9}\right)\epsilon,
\en
where we used the connection \eqref{connect} between the bulk modulus, the shear modulus and Poisson's ratio. To compare this stress with the simulated stress given in Figure \ref{fig:FEM0_5}, the volume change $\epsilon = i_3-1$ is needed. In Figure \ref{fig:shear-i_3}(b)  $i_3$, as measured by ABAQUS, is plotted as a function of the amount of shear $\gamma$ for different values of Poisson's ratio: $\nu =0.499, 0.495, 0.49$. 
In Table 1, we collected in particular results obtained when the amount of shear is $\gamma=1.0$, as in the simulations. 
The volume change  \color{black} $i_3-1$  \color{black}is displayed in the second column and the normalized Cauchy stress component $T_{22}/\mu$, found from \eqref{negative}, in the third column.
\begin{table} [h!]
\label{i3values}
\begin{center} 
\begin{tabular}{ccc} 
$\nu$ & \color{black}$i_3-1$ \color{black}& $T_{22}/\mu$\\ 
\hline 
0.499 & \color{black} 0.0009 \color{black}& 0.116 \\ 
0.495 &  \color{black} 0.0029  \color{black} & -0.045 \\ 
0.490 &  \color{black} 0.0044  \color{black} & -0.115 \\ 
\end{tabular} 
\caption{Variation of volume change with Poisson's ratio}
\end{center} 
\end{table}

It follows that \eqref{negative} predicts normal stresses that are very close to the average values of $T_{22}$ predicted by the numerical simulations given in Figure \ref{fig:gent}(b). Therefore infinitesimal changes in Poisson's ratio \textit{close to the limiting value of} $\nu =1/2$ may indeed result in dramatic changes in the value for the normal stress, even leading to an inverse Poynting effect, due to the volume changes which accompany every deformation.


\section{Conclusion}


It has been shown that the usual practice of arbitrarily choosing a value of Poisson's ratio when numerically modeling rubbers results in widely different normal stress distributions when the third decimal of $\nu \in [0.495,0.5]$ is modified. 
The sign of the Poynting effect may even be reversed.
It is thus essential for the accurate modeling of rubbers to experimentally determine Poisson's ratio to the utmost precision. 
This is not as hard as it seems, especially when the measure of $\nu$ relies on ultrasonic wave speeds evaluation. 
There, the contrast between the speed of a compression wave $v_L$ and the speed of a shear wave $v_T$ is so large for nearly incompressible solids, that the connection
\be \label{ultras}
\nu = \tfrac{1}{2}\dfrac{v_L^2-2v_T^2}{v_L^2-v_T^2},
\en 
gives excellent precision.
For instance, Wood and Martin \cite{WoMa64} find $\nu = 0.49986$ using acoustic waves.
Similarly, Gennisson et al. \cite{Genn07} measure the following body wave speeds for an Agar-Agar gel: $v_L = 1500$ m/s and $v_T = 1$ m/s,
giving $\nu = 0.49999978$, according to  \eqref{ultras}.

\color{black}
The issues raised here ought to be addressed by anyone wishing to use a commercial FE code for large deformation behavior.
For instance the Abaqus/Explicit manual \cite{abaq} indicates that the default setting for slight compressible models is $\nu = 0.475$, a value which allows volume changes sufficient to inverse the Poynting effect according to Figure 1. 
It goes on to state that if we ``are defining the compressibility rather than accepting the default value, an upper limit of 100 is suggested for the ratio of $\kappa/\mu$'', which corresponds to $\nu \simeq 0.495$, also within the range of extreme sensitivity in shear. 
Furthermore, the manual goes on to say that slight compressibility ``does not warrant special attention for plane stress'', a statement clearly at odds with our findings.
\color{black}


\section*{Acknowledgments}


This work was partly supported by a Marie Curie Fellowship for Career Development awarded by the Seventh Framework Programme (European Commission) to the first author.
We thank Aisling N\`i Annaidh for good pointers in the FE literature.



\end{document}